\documentstyle[prc,aps,manuscript]{revtex}

\newcommand{\Image}{\mathop{\rm Im}\nolimits}
\begin{document}

\title{E1 TRANSITIONS BETWEEN SPIN-DIPOLE AND
GAMOW-TELLER GIANT RESONANCES}

\author{V.A.~Rodin$^{1,3}$ and M.H.~Urin$^{2,3}$}
\address{$^1$ Kernfysisch Versneller Institute, 9747AA Groningen,
The Netherlands}
\address{$^2$
Research Center for Nuclear Physics, Osaka
University, Mihogaoka 10-1, Ibaraki, Osaka 657-0047,
Japan}
\address{$^3$ Moscow Engineering Physics Institute, 115409 Moscow, Russia}

\maketitle
\begin{abstract}
The branching ratios for E1 transitions between the spin-dipole (SD) and
Gamow-Teller (GT) giant resonances in $^{90}$Nb and
$^{208}$Pb are evaluated.
Assuming the main GT-state has the wave function close to that for the ``ideal"
GT-state, we reduced the problem to calculate the SD and GT strength
functions. These strength functions are evaluated
within an extended continuum-RPA approach.
\end{abstract}


{\bf 1.} Experimental and theoretical studies of direct-decay properties of
various
giant resonances (GRs) allow to check their
microscopic (particle-hole) structure in a quantitative way.
Experimentally the
partial branching ratios for the direct proton decay of the GTR
and SDR$^{(-)}$ are obtained from the
($^3${He},t) and ($^3${He},tp) experiments. The data at
E($^3${He})=450~MeV have been analyzed for the $^{208}${Pb}
target-nucleus~\cite{akimun,har} and have been rather successfully described
within
an extended continuum-RPA approach \cite{mouk}.
The data of the $^{90}$Zr($^3${He},tp) reaction are expected to be
analyzed soon~\cite{priv1}.
Another possibility to investigate the microscopic
structure of the SDR$^{(-)}$ and GTR is to study $\gamma$-transitions between
these
resonances. The branching ratios for the $\gamma$-decay from the SDR$^{(-)}$ to
the GTR can be deduced from the ($^3${He},t$\gamma$) coincidence experiments
~\cite{priv2}.

The intensity of the E1 $\gamma$-transitions between GT and SD$^{(-)}$
states in $^{90}$Nb and $^{48}$Sc was evaluated within
a TDA-approach in Ref.~\cite{suz}. However, the results obtained in this work
are presented in the form, which does not allow to compare them
directly with the
experimental branching ratios.  The aim of the present work is to evaluate the
branching ratio for the E1 transitions between the SDR$^{(-)}$ and the GTR (main
peak) in
$^{208}$Bi and $^{90}$Nb within the approach given in Ref.~\cite{mouk}.
In this approach we use:\\
(i) the continuum-RPA (CRPA);\\ (ii) the phenomenological mean field and the
Landau-Migdal
particle-hole interaction together with some partial self-consistency
conditions;\\ (iii) a phenomenological description of the doorway-state coupling
to many-quasiparticle configurations.

{\bf 2.}   We start from consideration of the CRPA polarizabilities
${\cal P}_{JLS}^{(-,+)}(\omega)$ and the strength
functions $S_{JLS}^{(-,+)}(\omega)$ corresponding to the external fields
$V_{JLSM}^{(-,+)}=\sum\limits_{a}V_{JLSM}^{(-,+)}(x_a)$.
Here,
$V_{JLSM}^{(-,+)}(x)=r^LT_{JLSM}
(\vec n)\tau ^{(-,+)}$ (with $J=S=1;L=0$ and
$J=0,1,2;L=S=1$ for GT and SD excitations, respectively), $T_{ {JLSM}}
(
\vec n)=\sum\limits_{m} C^{JM}_{Lm1M-m} Y_{Lm}(
\vec n) \sigma^{M-m} $ is the
irreducible spin-angular tensor operator of the rank $J$, and $\sigma^\mu$ and
$\sqrt{2} \tau^{(\pm)}$ are the spherical spin and isospin Pauli matrices,
respectively; $\omega$ is the excitation
energy measured from the energy of the parent-nucleus ground-state. For the
considered spin GRs, the CRPA polarizabilities and the strength functions
exhibit
resonance-like behaviour, corresponding to the excitation of isolated
particle-hole type doorway states.
In particular, using the Breit-Wigner parameterization
of the polarizabilities and the strength functions:
\begin{eqnarray}
S_{J11}^{(-)}(\omega)=-\frac1\pi\Image{\cal P}_{J11}^{(-)}(\omega)
=-\frac1\pi\Image\sum\limits_{s}\frac{R_{s}}{\omega-
\omega_{s}+\frac{\rm{i}}2\Gamma_{s}^\uparrow}\ ;
\label{eq_0}
\\
S_{101}^{(-)}(\omega)=-\frac1\pi\Image{\cal P}_{101}^{(-)}(\omega)
=-\frac1\pi\Image
\frac{R_g}{\omega-
\omega_g+\frac{\rm{i}}2\Gamma_g^\uparrow}\
\label{eq_1}
\end{eqnarray}
we can evaluate the doorway parameters:
strength $R_{s(g)}$, energy $\omega_{s(g)}$, and total escape width
$\Gamma^\uparrow_{s(g)}$. Similarly to the work of Ref.~\cite{mouk}, we consider
only the main GT doorway-state with the maximal strength $R_{g}$.

{\bf 3.} The radiative  width
 for the E1-transitions between SD$^{(-)}$ and GT
doorway-states is determined by the squared matrix elements of the electric
dipole
operator $D_{\mu}^{(3)}=-\frac12 e
\sum\limits_a r_a Y_{1\mu}(
\vec n_a)\tau_a^{(3)}$ according to the expression:
\begin{equation}
\Gamma^\gamma_{s\to g}=\displaystyle\frac{16\pi}{9}
\left(\displaystyle\frac{\omega_s-\omega_g}{\hbar c}\right)^3
\sum_\mu{\overline{\left|\left(D_\mu^{(3)}\right)_{gs}\right|^2}}.
\label{eq1}
\end{equation}
Here, the bar denotes averaging over $M_s$ and summation
over $M_g$, where $M$ are projections of the doorway-state total angular
momentum. To describe the radiative width $\Gamma^\gamma_{s\to g}$ in terms of
the SD$^{(-)}$ doorway-state strength
$R_{s}$, we start from the assumption that
the component of main GT state with projection $M_G$ exhausts the total
GT strength $R_{G}=\displaystyle\frac{(N-Z)}{4\pi}$. With this assumption
the following equations,
which are similar to those used in Ref.~\cite{suz}, are valid:
\begin{eqnarray}
|G\rangle=R^{-1/2}_{G} V^{(-)}_{101M_{G}}|0\rangle\ , \
 V^{(+)}_{101M_{G}}|0\rangle=0\ , \nonumber\\
\left(D_\mu^{(3)}\right)_{sG}=R^{-1/2}_{G}
\left(\left[D_\mu^{(3)},V^{(-)}_{101M_{G}}\right]\right)_{s0}.
\label{eq2}
\end{eqnarray}
Here, $|G\rangle$ is the wave function of the ``ideal" GTR, $|0\rangle$ is the
parent-nucleus ground-state wave function.
Note that the commutator in eq.(\ref{eq2}) is a component
of the operator $V^{(-)}_{J11M}$ divided by $\sqrt{4\pi}$.
Assuming the main GT
state has the wave function close to that for the ``ideal" GT state, we can
derive from
eqs.~(\ref{eq1}) and (\ref{eq2}) the expression for the radiative width
$\Gamma^\gamma_{s\to g}$:
\begin{equation}
\Gamma^\gamma_{s\to g}=\frac{4 e^2}{9R_{G}}
\left(\displaystyle\frac{\omega_{s}-\omega_{g}}{\hbar c}\right)^3
x_{g}R_{s},
\label{eq4}
\end{equation}
where the factor $x_{g}=R_{g}/R_{G}$ is the strength of the main GT state
related to the total one.
Via this factor we take into account the difference between the ``ideal'' GT
state and
the main GT doorway state. The width
$\Gamma^\gamma_{S\to g}$ for the E1 transitions from the SDR$^{(-)}$
to the GTR (main peak) can be schematically described with the use of
eq.(\ref{eq4}).
Assuming all the $J$-components of the SDR$^{(-)}$ have
the same energy $\omega_{S}$, equal to the experimental SDR$^{(-)}$ energy,
and both GRs have no spreading and escape widths, the radiative width can
be expressed in terms of the non-energy-weighted sum rule for
spin-dipole transitions:
\begin{eqnarray}
&\Gamma^\gamma_{S\to g}=\displaystyle\frac{4 e^2}{9}
\left(\displaystyle\frac{\omega_{s}-\omega_{g}}{\hbar c}\right)^3
x_g\displaystyle\frac{\left<r^2\right>^{(-)}}{(1-B)}, \label{eq5} \\
& \left<r^2\right>^{(-)}=\displaystyle\frac{4\pi}{N-Z}\int
\varrho^{(-)}(r)r^4dr\ , \ \ \
B=\displaystyle\frac{R^{(+)}}{R^{(-)}}. \nonumber
\end{eqnarray}
Here, $B$ is the SDR$^{(+)}$ excitation strength related to that for the
SDR$^{(-)}$, and $\varrho^{(-)}(r)$ is the neutron-excess density.

{\bf 4.} The SD$^{(-)}$ strength distribution and the doorway-state coupling
to
many-quasiparticle configurations are  taken into account within the approach of
Ref.~\cite{mouk}.
Similarly to the SD$^{(-)}$ polarizability (to the ``forward-scattering
amplitude") of
eq.(\ref{eq_0}) we can also use the Breit-Wigner parametrization for the
``reaction
amplitude" $M^{J}_{g}(\omega)$, corresponding to both the excitation of
$J^{-}$ doorway states and their E1 decay to the main GT state:
\begin{equation}
M_g^J(\omega)=\displaystyle\frac 1{\sqrt{2\pi }}
\sum\limits_{s}\displaystyle\frac{R_{s}^{1/2}(\Gamma^\gamma_{s\to g})^{1/2}}
{\omega -\omega _{s}+\frac{\rm{i}}2
\Gamma_{s}^\uparrow}=\displaystyle\frac{\alpha^{1/2}_g}{\sqrt{2\pi }}
\sum\limits_{s}\displaystyle\frac{R_{s}\sqrt{(\omega_s - \omega_g)^{3}}}
{\omega -\omega_{s}+\frac{\rm{i}}2\Gamma_{s}^\uparrow}\ ,
\label{eq7}
\end{equation}
where $\Gamma^\gamma_{s\to g}$ is the radiative width of eq.~(\ref{eq4}) and
$\alpha_g = \displaystyle\frac{4 e^2}{9 (\hbar c)^3}
\displaystyle\frac{x_g}{R_G}$. Thus, the structure of
the resultive amplitude $M^{J}_g$ is found to be close
to that of the SD polarizability of eq.~(\ref{eq_0}).

Then the doorway-state coupling to many-quasiparticle configurations is
phenomenologically taken into account. To get the expressions for the
energy-averaged ``reaction amplitudes" we substitute the escape widths
$\Gamma_{s}^\uparrow$ in eqs.~(\ref{eq_0}), (\ref{eq7}) and the
width $\Gamma_{g}^\uparrow$ in eq.~(\ref{eq_1})
by $\Gamma_{s}^\uparrow+\Gamma_{S}^\downarrow$
and $\Gamma_{g}^\uparrow+\Gamma_{G}^\downarrow$, respectively.
The mean doorway-state spreading width $\Gamma_{S}^\downarrow$ is found
from the condition that the total width $\Gamma$ of the
SD$^{(-)}$ energy-averaged strength function
$\bar S_{SD}^{(-)}(\omega)=\sum\limits_{J=0,1,2}(2J+1)\bar
S^{(-)}_{J11}(\omega)$
coincides with the total width $\Gamma^{exp}$ of the SDR$^{(-)}$ in
the experimental inclusive reaction cross section. The same procedure is used
to evaluate $\Gamma_{G}^\downarrow$. Because the doorway-state spreading widths
$\Gamma_{S}^\downarrow$ and $\Gamma_{G}^\downarrow$ are found to be
rather large, we take approximately into account a
variation of factor E$_\gamma^3$ over the doorway-state resonances,
using in the expression for the squared
energy-averaged ``reaction amplitude" $\bar M_J(\omega)$
the corresponding averaged value:
\begin{equation}
\overline{(\omega_s-\omega_{g})^3}=(\omega_{s}-\omega_{g})^3
+3(\omega_{s}-\omega_{g})\sigma^2_{gs}\ , \
\sigma_{gs}=\sqrt{(\Gamma_{s}^\uparrow+\Gamma^\downarrow_{S})^2
+(\Gamma_{g}^\uparrow+\Gamma^\downarrow_{G})^2}/2.35.
\label{eq8}
\end{equation}
The ratio of the integrated energy-averaged ``cross sections":
\begin{equation}
b_{g} =\left. \int \sum\limits_{J=0,1,2}(2J+1)\left|\bar
M_{g}^{J}(\omega)\right|^2{\rm d}\omega \right/ \int \bar
S^{(-)}_{SD}(\omega){\rm d}\omega \ ,
\label{eq11}
\end{equation}
can be considered as the partial branching ratio for the E1-decay from
the SDR$^{(-)}$ to the GTR (main peak).
The branching ratio described schematically is
determined by using the width $\Gamma^\gamma_{gS}$ of eq.~(\ref{eq5}) as
$b^{schem}_g = \Gamma^\gamma_{gS}/\Gamma^\downarrow_S$.

{\bf 5.} The partial self-consistency conditions and choice of model parameters
are described in Ref.\cite{mouk}. In particular, the isoscalar mean field
amplitude $U_0$ and
the amplitude $f'$ of the isovector part of the Landau-Migdal
particle-hole interaction are chosen for each nucleus
to reproduce in calculations
the experimental proton and neutron separation energies.
The values of $U_0$ and $f'$ are listed in Table 1.
The ability of the model to describe the
single-neutron-hole spectrum of
$^{207}${Pb} has been demonstrated in Ref.~\cite{mouk}.

The spin-dipole sum rule $\int \varrho^{(-)}(r)r^4dr$
is  evaluated for $^{90}${Zr} and $^{208}${Pb} in the same way as described
in Ref.~\cite{mouk}. The sum rule is
determined by the mean-squared radius $\left<r^2\right>^{(-)}$ of
eq.~(\ref{eq5}) (the corresponding calculated values are listed
in Table 1).
The amplitude $g'$ of the spin-isospin part of the Landau-Migdal particle-hole
interaction is chosen for each nucleus to reproduce in calculations the
experimental GTR energy. The values of $g'$ are listed in Table 1 along with
the relative strengths of the GT main peak $x_{g}$.

The basic assumption used in this work is the substitution of the RPA creation
operator, corresponding to the main GT state, by the operator
$R^{-1/2}_{G} V^{(-)}_{101M_{G}}$ with taking $x_g$ as the correction factor
(see eqs.(\ref{eq2}),(\ref{eq4})). Although the accuracy of this assumption is
not-too-high because the $x_g$ values are not-too-close to unity
(Table 1), it seems enough to make a reasonable comparison
of the calculated branching ratios with coming experimental data.

The partial branching ratios $b_{g}^{schem}$ for the
gamma-decay of the SDR$^{(-)}$ to GTR are calculated within the
framework of the schematic description. In the calculations, the experimental
energy $21.1$ MeV \cite{akimun} ($17.9$ MeV \cite{goodman})
is used for the SDR$^{(-)}$ in $^{208}${Bi} ($^{90}${Nb})
along with the experimental energy $15.5$ MeV \cite{akimun}
($8.7$ MeV \cite{goodman}) for the GTR. The calculated values of
$b_{g}^{schem}$ are given in Table 1.

To take into account the distribution of the spin-dipole
particle-hole strength over
the SDR$^{(-)}$, the doorway-state spreading widths for both the SDR$^{(-)}$ and
GTR, we calculated partial radiative branching ratios $b_{g}$ by
eqs.~(\ref{eq7})-(\ref{eq11}) within the framework of the more refined
description. Along with the spin-dipole doorway-state parameters calculated in
the same way given in Ref.~\cite{mouk}, we used in the
calculations for $^{208}${Bi} ($^{90}${Nb}) \
(i) the mean spin-dipole doorway-state spreading width
$\Gamma_{SD}^\downarrow=4.7$ MeV found in Ref.~\cite{mouk}
($\Gamma_{SD}^\downarrow=5.0$ MeV to reproduce
the experimental total SDR width
$\Gamma_{SDR}^{exp}=7.8$ MeV \cite{goodman})
and (ii) the experimental total width of the GTR
$\Gamma_{GTR}^{exp}=(\Gamma_{GTR}^\downarrow+\Gamma_{GTR}^\uparrow)^{exp}=
3.72$ MeV \cite{akimun} ($\Gamma_{GTR}^{exp}=4.4$ MeV \cite{goodman}).
The calculated values of $b_{g}$ are given in Table 1.

The partial branching ratios $b_{g}^{schem}$ and $b_{g}$ for the $\gamma$-decay
from the SDR$^{(-)}$ to the GTR
in  $^{208}${Pb} are rather different (Table 1). The difference is mainly due to
variation of the factor E$_\gamma^3$ over the SDR$^{(-)}$
doorway states taken into account within the realistic description.
It is also noteworthy that the calculated mean SDR$^{(-)}$-energy
$23.1$ MeV \cite{mouk} is higher than the experimental value of
$21.1\pm0.8$ MeV \cite{akimun} (for $^{90}${Nb} the corresponding values are
$19.2$ MeV and $17.9$ MeV \cite{goodman}, respectively).

In conclusion, we evaluate the branching ratios of the $\gamma$-decay from
the SDR$^{(-)}$ to the GTR in
$^{208}${Bi} and $^{90}${Nb} within the extended continuum-RPA
approach. These predictions are expected to be appropriate for a comparison
with the corresponding experimental data.

{\bf 6.} The authors are grateful to M.N. Harakeh, M. Fujiwara, and A.
Krasznahorkay for stimulating discussions and valuable remarks.
The authors acknowledge generous financial supports from the
``Nederlansdse organisatie voor wetenshappelijk onderzoek" (V.A.R.)
and from the RCNP COE fund (M.H.U.).

\begin{table}
\caption{ The calculated branching ratios for the $\gamma$-decay from the
SDR$^{(-)}$ to the GTR (main peak) in $^{90}$Zr and $^{208}$Pb.
The mean-squared radii calculated by
eq.(6) and the $R^{(+)}$ to $R^{(-)}$ ratios $B$ are also
given together with the isoscalar mean field amplitude $U_0$, the Landau-Migdal
parameters $f'$ and $g'$, and the calculated relative strengths $x_{g}$ of the
GT main peak.}
\begin{center}
\begin{tabular}{|c|c|c|c|c|c|c|c|c|}
\hline
Nucleus & $U_0,$ MeV& $f'$&$g'$&$x_g$& $\left<r^2\right>^{(-)}$, fm$^2$
& $B$& $b_{g}^{schem}\: (\times 10^{-4})$ & $b_{g}\: (\times 10^{-4})$ \\
\hline
$^{90}$Zr &53.3&0.96& 0.70&0.83&22.6 & 0.34& 3.3& 4.7 \\
\hline
$^{208}$Pb &54.1&1.0& 0.78&0.69&36.4 & 0.06& 0.81& 2.4 \\
\hline
\end{tabular}
\end{center}
\end{table}


\end{document}